\documentstyle[epsf]{aipproc}
\begin{document}
\title{An eigenfunction method for particle acceleration at ultra-relativistic shocks}

\author{Axel W. Guthmann$^*$, John G. Kirk$^*$,
Yves A. Gallant${^{\dagger, \ddagger}}$ and Abraham Achterberg$^{\ddagger}$}
\address{$^*$Max-Planck-Institut f\"ur Kernphysik, Postfach 103980 ,
D-69029 Heidelberg, Germany  \thanks{homepage: http://www.mpi-hd.mpg.de/theory/}\\
$^{\dagger}$ Astronomical Institute, Utrecht University, P.O. Box 80000, 3508 TA Utrecht,
 Netherlands\\
$^{\ddagger}$ Dublin Institute for Advanced Studies, 5 Merrion Square, Dublin 2, Ireland\\}
\maketitle

\begin{abstract}
We adapt and modify the eigenfunction method of computing the power-law spectrum
of particles accelerated at a relativistic shock front via the
first-order Fermi process
\cite{aguthmann:kirkschneider87}
 to apply to shocks of arbitrarily high Lorentz
factor. The power-law index of accelerated particles undergoing
isotropic small-angle scattering at an ultrarelativistic, unmagnetized 
shock is found to be $s=4.23\pm0.2$ (where $s=d\ln f/ d\ln p$,                      
with $f$ the Lorentz-invariant phase-space density and $p$ the
momentum), in agreement with the results of Monte-Carlo simulations. 
We present results for
shocks in plasmas with different equations of state and for Lorentz 
factors ranging from 5 to infinity. 

\end{abstract}

\section*{The method}
We study a stationary shock front in the $x-y-$plane. The accelerated
particles are assumed to be test-particles without influence on the
dynamics of the plasma or the jump conditions at the
shock-front. The plasma flows
along the $z$-axis, with constant velocities $u_{-}$ 
in the upstream ($z<0$) region and $u_{+}$ downstream ($z>0$), the 
velocities are related by the Rankine-Hugoniot jump conditions.

Test-particles are injected into the acceleration process and their interaction with the
plasma flow is assumed to give rise to diffusion in the angle $\cos^{-1}\mu$ between a particle's
 velocity and the shock normal. In the frame of the shock front this leads to
a stationary transport equation valid for the local plasma rest frame and given in mixed coordinates
as \cite{aguthmann:kirkschneider87}
\begin{equation}\label{E:aguthmann:1}
   \Gamma(u+\mu)\frac{\partial f}{\partial
   z}=\frac{\partial}{\partial \mu}
   D_{\mu \mu}(1-\mu^{2})\frac{\partial f}{\partial \mu}
\end{equation}
where the plasma speed $u$ is measured in units of the speed of light,
$\Gamma=\left(1-u^{2}\right)^{-1/2}$ is the Lorentz-factor,
$f(p,\mu,z)$ is the (Lorentz invariant)
phase-space density as a function of the particle 
momentum $p$, direction $\mu$
and position. $p$ and $\mu$ are measured in the
local rest frame of the plasma, whereas $z$ is measured in the rest frame of 
the shock front.

Equation (\ref{E:aguthmann:1}) is
solved using the separation {\it Ansatz}  
 \cite{aguthmann:kirkschneider87}
\begin{equation}\label{E:aguthmann:2} 
  f(p,u,\mu,x)=\sum_{i=-\infty}^{+\infty}g_i(p)Q_i(\mu,u)\exp\left(\Lambda_iz
/\Gamma\right),
\end{equation}
valid in each half-plane  
with $\Lambda_i$ and $Q_i$ the eigenvalues and
eigenfunctions of the equation 
\begin{equation}
\label{E:aguthmann:3}
\left\{ \frac{\partial}{\partial \mu} \left[
  D_{\mu\mu}\frac{\partial}{\partial \mu} \right]-\Lambda_i(u+\mu) \right\} 
Q_i(\mu,u)=0 
\end{equation}
The momentum distribution of particles with energy far above the injection energy range 
-- those in which we are interested -- takes the shape of a power-law $g_i(p)\propto p^{-s}$ with a power-law index $s$,
since there is no preferred  momentum scale in this range.

Matching the expansion (\ref{E:aguthmann:2}) across the shock front 
according to Liouville's Theorem and
imposing physically realistic boundary conditions up and downstream
leads to a nonlinear algebraic equation for the power law index $s$.

\begin{figure}
\centerline{\epsfxsize=7.5cm\epsffile{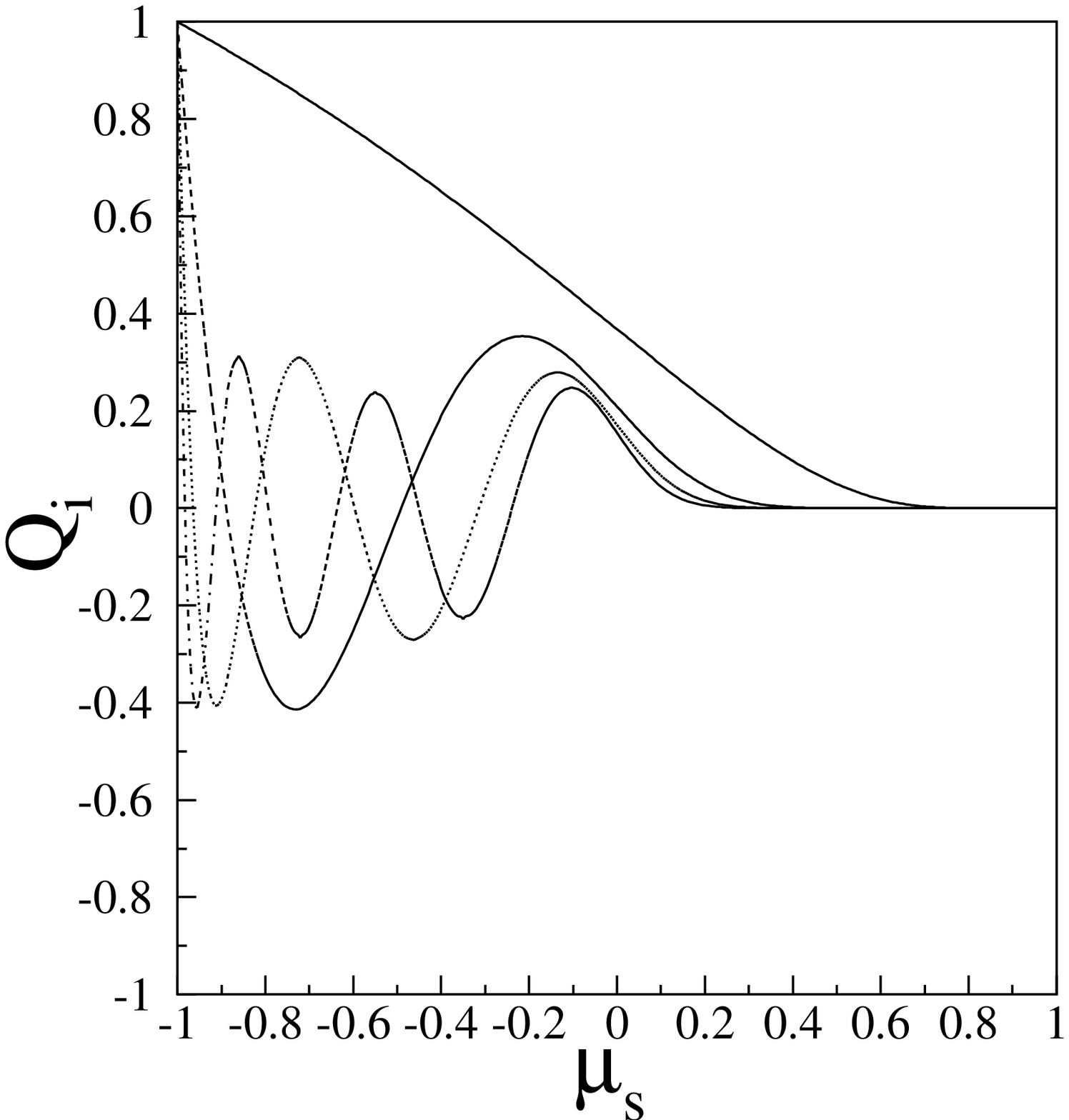}\epsfxsize=7.5 cm\epsffile{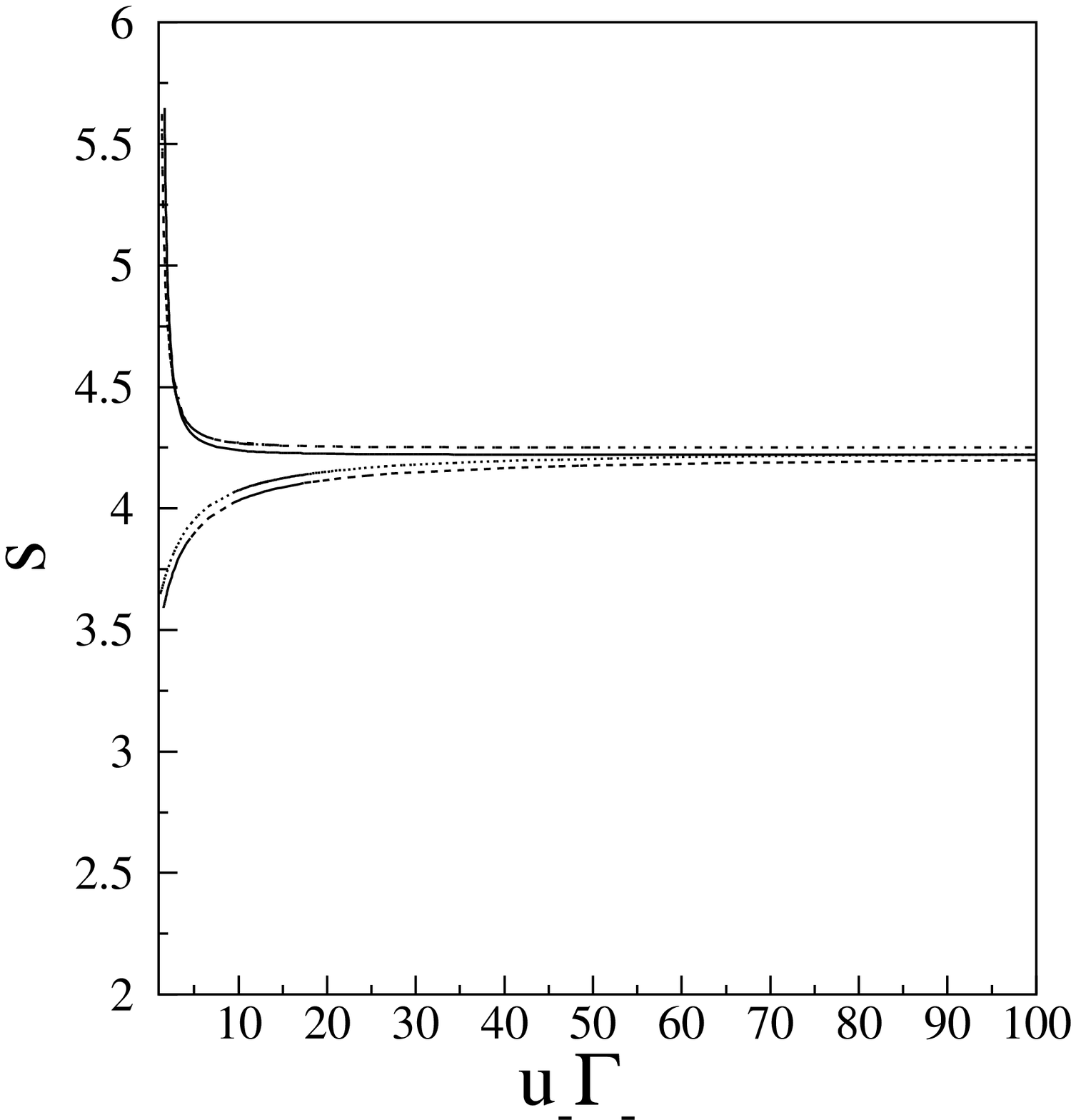}}
\caption{A) (left) The eigenfunctions $Q_i$ for $i=1,3,5,7$ for $\Gamma_-=223$,
as a function of the 
(cosine of the) angle between the particle speed and the 
shock normal, measured
in the shock frame, for a relativistic gas.
B) (right) The power-law index $s$ for relativistic gas with isotropic 
(solid, 2nd from top) and anisotropic 
(dashed-dotted, top) scattering operator and 
for a strong shock in a gas of adiabatic index $4/3$, with 
isotropic (dashed, 4th from top) and 
anisotropic (dotted, 3rd from top) scattering
}
\protect\label{f:aguthmann:1}
\end{figure}

In \cite{aguthmann:kirkschneider87} and \cite{aguthmann:heavensdrury88}  
only the eigenfunctions with $i<0$ were used and the method 
was applied to mildly relativistic shock speeds ($\Gamma_-\le5$).
Here, we use the eigenfunctions with $i>0$ and 
calculate them directly with a 
numerical scheme. In the limit $u_-\rightarrow1$ an analytic 
expression is available \cite{aguthmann:kirkschneider89}. 
Four eigenfunctions ($i=1,3,5,7$) are shown in Fig.~\ref{f:aguthmann:1}A 
as functions of the cosine $\mu_{\rm s}=(\mu+u)/(1+\mu u)$ 
of the angle between the particle direction and the shock normal, 
measured in the shock rest frame.
For $i>1$ they are oscillatory for $-1 < \mu_{\rm s} < 0$ and for all $i>0$ 
fall off monotonically in the
range $0<\mu_{\rm s}<1$.

\section*{Results}
The index $s$ of the momentum spectra of the 
accelerated particles in different cases are shown in 
Fig.~\ref{f:aguthmann:1}B. The jump conditions investigated are those 
for a relativistic gas both up and downstream:
$u_-u_+=1/3$\label{S:aguthmann:1} and
for a strong shock in a medium with adiabatic index $4/3$ \cite{aguthmann:kirkduffy99}.
 
Also we investigate two 
different scattering operators, $D_{\mu\mu 1}=1-\mu^2$ (isotropic small/angle
scattering)
and $D_{\mu\mu 2}=(1-\mu^2)\times(\mu^2+0.01)^{1/3}$
corresponding to scattering in weak Kolmogorov turbulence, together with 
a rough prescription for avoiding the lack of scattering at $\mu=0$ \cite{aguthmann:heavensdrury88}. 
For high upstream Lorentz-factors 
the power-law index settles at a value around $4.23$ for all equations of state,
which is reproduced in the limiting case $u_-\rightarrow1$. 
The scattering operator  has only a minor effect.

\section*{Summary}
These results are in agreement with  
the asymptotic Monte-Carlo results of Gallant et al.
\cite{aguthmann:gallant} and those of Bednarz \& Ostrowski
\cite{aguthmann:bednarz} for $\Gamma_-\approx200$.
Anisotropic scattering, which has not been treated by Monte-Carlo simulations, 
leads to a slight steeping in the power-law spectrum,
 because fewer particles are able to cross the region $\mu \approx 0$
 and return to the shock.
From observations of GRB afterglows, Galama et al. 
\cite{aguthmann:galama} and Waxman \cite{aguthmann:waxmann}
have found synchrotron spectral indices 
corresponding to $s\approx4.25$, 
implying that the particles could 
indeed have been accelerated by the first order Fermi mechanism operating at
an ultrarelativistic shock front.

\subsection*{ }
This work was supported by the European Commission under the TMR programme, contract number ERBFMRX-CT98-0168

\end{document}